# TRANSITS OF EARTH-LIKE PLANETS

LISA KALTENEGGER

Harvard-Smithsonian Center for Astrophysics, 60 Garden Street, Cambridge, MA 02138, USA; lkaltenegger@cfa.harvard.edu

WESLEY A. TRAUB

Jet Propulsion Laboratory, M/S 301-451, 4800 Oak Grove Drive, Pasadena, CA 91109; also Harvard-Smithsonian Center for Astrophysics

## ABSTRACT

Transmission spectroscopy of Earth-like exoplanets is a potential tool for habitability screening. Transiting planets are present-day "Rosetta Stones" for understanding extrasolar planets because they offer the possibility to characterize giant planet atmospheres and should provide an access to biomarkers in the atmospheres of Earth-like exoplanets, once they are detected. Using the Earth itself as a proxy we show the potential and limits of the transiting technique to detect biomarkers on an Earth-analog exoplanet in transit. We quantify the Earth's cross section as a function of wavelength, and show the effect of each atmospheric species, aerosol, and Rayleigh scattering. Clouds do not significantly affect this picture because the opacity of the lower atmosphere from aerosol and Rayleigh losses dominates over cloud losses. We calculate the optimum signal-to-noise ratio for spectral features in the primary eclipse spectrum of an Earth-like exoplanet around a Sun-like star and also M stars, for a 6.5-m telescope in space. We find that the signal to noise values for all important spectral features are on the order of unity or less per transit - except for the closest stars - making it difficult to detect such features in one single transit, and implying that co-adding of many transits will be essential.

*Subject headings*: occultation, Earth, astrobiology, eclipse, atmospheric effects, techniques: spectroscopic

## 1. INTRODUCTION

The transiting system CoRoT-7b (Leger et al. in prep) shows that transiting Super-Earths have already been detected, and recent detections of several Super-Earths (Mayor et al. 2009) show that transiting Earths are expected to be detected in the near future. The current status of exoplanet characterization shows a surprisingly diverse set of planets. For a subset of these, some properties have been measured or inferred using radial velocity, micro-lensing, transits, and astrometry. These observations have yielded measurements of planetary mass, orbital elements, planetary radius (for transits), and some physical characteristics of the upper atmospheres. Specifically, observations of transits, combined with radial velocity (RV) information, have provided estimates of the mass and radius of the planet (see e.g., Torres et al. 2008), planetary brightness temperature (Charbonneau et al. 2005; Deming et al. 2005), planetary day-night temperature difference (Harrington et al. 2006; Knutson et al. 2007), and even absorption features of planetary upper-atmospheric constituents: sodium (Charbonneau et al.

2002), hydrogen (Vidal-Madjar et al. 2003), water (Tinetti et al. 2007,(disputed by Ehrenreich et al. 2007), Beaulieu et al 2008, Swain et al. 2008) and methane (Swain et al. 2008), showing that the transit technique has great value. Several groups are modeling the transmission spectra of extrasolar giant planets in detail (see e.g. the review article by Charbonneau et al. 2006 and references therein). That success has led to speculation that the transit technique might also be useful for characterizing terrestrial planets.

In this paper we use the Earth itself as a proxy to show the potential, and limits, of the transiting technique to detect biomarkers on Earth-analog exoplanets. We calculate the visible and infrared transit spectra of the Earth. With this information we calculate the signal to noise ratio (SNR) for major spectral features, for the case of a 6.5-m telescope in space, like JWST, during the time of a single transit, as well as for co-added transits, for a Sun-like star and for M stars. We note that M stars have been suggested as good targets for characterizing a planet's atmosphere with transmission spectroscopy due to the improved contrast ratio between star and planet. Ground based transit searches are underway focusing on M stars (see e.g. Irwin et al. 2009).

Theoretical transmission spectra of terrestrial exoplanets have been published by Ehrenreich et al. (2006) in the wavelength range from 0.2-2 μm for simplified atmospheric profiles consisting of water vapor ($H_2O$), molecular oxygen ($O_2$), ozone ($O_3$), carbon dioxide ($CO_2$) and molecular nitrogen ($N_2$), and an opaque cloud layer below 10 km for a F2, K2 and G2 dwarf star. The work presented in this paper extends the wavelength range of the calculations to the infrared 0.3-20 μm, uses a realistic atmospheric temperature profile with aerosol, Rayleigh scattering, and three different cloud layers, validates the model with ATMOS 3 infrared transmission spectra of the Earth's limb, and presents a complete set of SNR calculations of atmospheric species during the transit for the Sun as well as M0 to M9 dwarf stars.

Currently fifteen exoplanets (including three pulsar planets) are known to have a mass (times sin i, where i is the orbital inclination, for RV planets) less than 10 $M_{Earth}$, a somewhat arbitrary boundary that distinguishes terrestrial from giant planets (Valencia et al. 2006 and references therein). Accordingly, we identify masses in the range 1-10 $M_{Earth}$ as being Super Earths, likely composed of rock, ice, and liquid, and masses greater that 10 $M_{Earth}$ as being giant planets, likely dominated by the mass of a gaseous envelope. The fifteen planets are: COROT-7b ∼7 $M_{Earth}$ (Leger et al. in prep), GJ 876 d, ∼7.5 $M_{Earth}$ (Rivera et al. 2005); OGLE-05-390L b, ∼5.5 $M_{Earth}$ (Beaulieu et al. 2006); Gl 581 c and Gl 581 d, ∼ 5.03 $M_{Earth}$ and 8.6 $M_{Earth}$ (Udry et al. 2007); HD40307 b, HD40307 c and HD40307 d∼4.2, 6.7, and 9.4 $M_{Earth}$ (Mayor et al. 2009); MOA-2007-BLG-192L b ∼3.3 $M_{Earth}$ (Bennett et al. 2008); HD 181433 b ∼7.6 $M_{Earth}$ (Bouchy et al. 2009); HD 285968 b ∼8.4 $M_{Earth}$ (Forveille et al. 2009); HD 7924 b ∼9.2 $M_{Earth}$ (Howard et al. 2009); as well as three planets discovered by pulsar timing (Wolszczan & Frail 1992). None of those planets orbits its star within the habitable zone (HZ), but Gl581 c and especially Gl581 d are close to the HZ edges (see Selsis et al. 2007, Kaltenegger et al. in prep).

In this paper we ask, what are the limits to characterizing an Earth-analog during a transit? We explore potential spectral signatures and biomarkers for an Earth-like planet. This translates into the accuracy needed to measure the planet's effective radius to detect atmospheric species. We calculate a model transmission spectrum from the UV to mid-IR, including realistic opacities and clouds, and use this to assess how well we could characterize our own planet in transit.

2. MODEL DESCRIPTION

Model Earth spectra are calculated with the Smithsonian Astrophysical Observatory code originally developed to analyze balloon-borne far-infrared thermal emission spectra of the stratosphere, and later extended to include visible reflection spectra (Traub & Stier, 1976; Jucks et al. 1998; Traub & Jucks, 2002). The spectral line database includes the large HITRAN compilation plus improvements from pre-release material and other sources (Rothman et al. 2004; 2009; Yung & DeMore 1999). The far wings of pressure-broadened lines can be non-Lorentzian at around 1,000 times the line width and beyond; therefore, in some cases ($H_2O$, $CO_2$, $N_2$) we replace line-by-line calculation with measured continuum data in these regions. Aerosol and Rayleigh scattering are approximated by applying empirical wavelength power laws (Allen 1976, Cox 2000) that exert an appreciable effect in the visible blue wavelength range. Atmospheres from 0 to 100 km altitude are constructed from standard models that are discretized to appropriate atmospheric layers. We do line-by-line radiative transfer through the refracting layers of the atmosphere. Clouds are represented by inserting continuum-absorbing/emitting layers at appropriate altitudes, and broken clouds are represented by a weighted sum of spectra using different cloud layers. For the transit spectra, we assume that the light paths through the atmosphere can be approximated by incident parallel rays, bent by refraction as they pass through the atmosphere, and either transmitted or lost from the beam by single scattering or absorption. We model the Earth's spectrum using its spectroscopically most significant molecules, $H_2O$, $O_3$, $O_2$, $CH_4$, $CO_2$, CFC-11, CFC-12, $NO_2$, $HNO_3$, $N_2$ and $N_2O$, where $N_2$ is included for its role as a Rayleigh scattering species. For this paper we use the US Standard Atmosphere 1976 spring-fall pressure-temperature profile, (COESA 1976, Cox 2000) and mixing ratio profiles shown in Fig.1. For further details on the model in an exoplanet context see Turnbull et al. (2006) and Kaltenegger et al. (2007).

We divide the atmosphere into 30 thin layers from 0-100 km altitude with thinner layers closer to the ground. The spectrum is calculated at very high spectral resolution, with several points per line width, and often thousands of points in the wings. The line shapes and widths are computed using Doppler and pressure broadening on a line-by-line basis.

The Rayleigh scattering optical depth $\tau_R(\sigma)$, at wavenumber $\sigma$ (cm$^{-1}$), in each discrete layer, is given by

$$\tau_R(\lambda) \cong 4.065 \times 10^{-44} \sigma^4 N \qquad (1)$$

where $\sigma = c/\lambda$, $c$ is the speed of light, $\lambda$ is wavelength, and $N$ is the column density of air (molecules cm$^{-2}$) along the ray path in the layer. Scattered photons are treated as lost to absorption, since we do not include multiple scattering in this code. We ignore a correction factor to this formula that reflects the increased polarisability of molecules at short wavelengths, which would increase the optical depth by about 1 %, 5 %, and 14 % at 1.0 μm, 0.5 μm, and 0.3 μm, respectively, and which we ignore for computational speed.

$$\tau_A(\lambda) \cong 8.85 \times 10^{-33} \sigma^{1.3} N \qquad (2)$$

This actual optical depth at low altitudes may well be larger than this value, under hazy or dusty conditions, according to Allen, however at such altitudes it is likely that clouds will already be blocking the light path. Also, the Rayleigh optical depth is larger than the aerosol depth for wavelengths less than 0.63 μm, therefore Rayleigh dominates in that region, and at longer wavelengths both are relatively small, about 0.05 per atmosphere and less.

### 2.1. Primary Transit

In this section we calculate the SNR for the detection of strong spectral features on an Earth-twin planet as it crosses the disk of its star, i.e., a transit. Here "Earth-twin" means a planet with the same mass, radius, temperature, and atmosphere as the Earth. A super-Earth will give essentially identical results, as discussed in Kaltenegger & Traub (in prep.), because the atmosphere's annulus area will remain to first order constant as the mass of a rocky planet increases - ignoring the compressibility of the solid body. We use a simple geometrical model in which the atmosphere of a spherical Earth is modeled with a layered shell geometry. The overall high-resolution spectrum is calculated, and smeared to lower resolution. To calculate the transmission spectrum of the full atmosphere, we first calculate the absorption spectrum for 30 probe rays that have tangent heights in the range 0-100 km. Each ray is tangent to an annulus of the atmosphere and therefore crosses several higher layers of the atmosphere along the line of sight, as it enters and exits the atmosphere, along its slightly refracted path. We then model the area of the stellar disk that is blocked by a

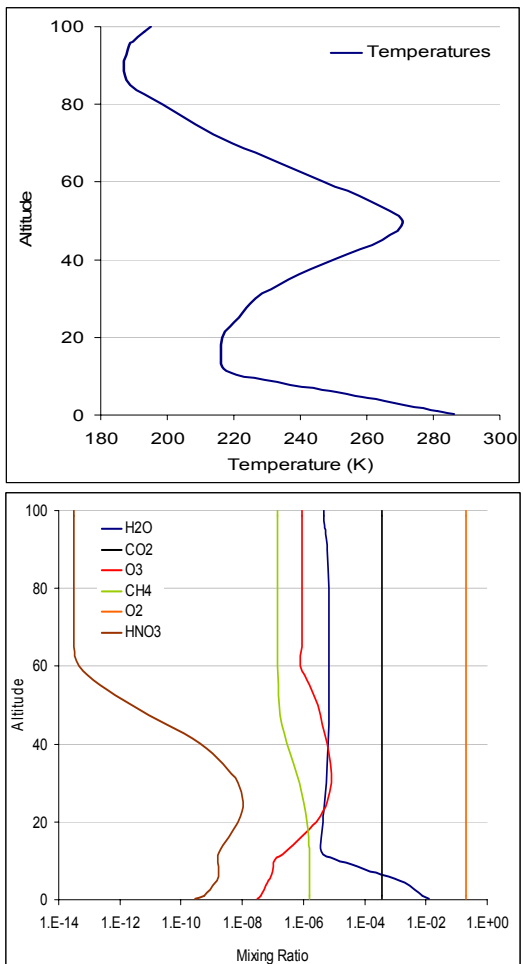

**Fig.1: (left)** Temperature profile, from US Standard Atmosphere 1976 (spring/autumn). **(right)** Mixing ratios versus height for the major detectable atmospheric gases up to 100 km (COESA 1976)

The aerosol absorption optical depth $\tau_A(\sigma)$ is included in a similar way. We assume that aerosols are uniformly mixed in the atmosphere, and assign a corresponding optical depth on a per-molecule basis, similar to the treatment of Rayleigh scattering. We use the magnitude and wavelength dependence of the aerosol optical depth in Allen (1976), which is appropriate for "very good conditions":

transiting planet as $\pi R^2(\lambda)$ where $R(\lambda) = R_p + h(\lambda)$, where $R_p$ is the radius of the planet at the base of the atmosphere, and $h(\lambda)$ is the effectively opaque height of the atmosphere at that wavelength. For example, $h(\lambda) = 0$ at wavelengths where the atmosphere is perfectly transparent, and $h(\lambda)$ can be as large as about 50 km for strong spectral features in the Earth's atmosphere.

The value of the effective height $h(\lambda)$ is calculated as follows. The fractional transmission $T_i(\lambda)$ is calculated for the i-th incident ray that passes through the atmosphere at a tangent height $h_i$, as discussed above. The next higher ray has a tangent height $h_{i+1} = h_i + \Delta h_i$. Then assuming that the atmosphere is a thin shell atop a large planet, the effective height $h(\lambda)$ is approximately given by

$$h(\lambda) = \int A(z,\lambda)\, dz = \Sigma\, A_i(\lambda)\, \Delta h_i \qquad (3)$$

where $A_i = 1 - T_i$ is the absorption along the i-th ray. The effective height $h(\lambda)$ is the altitude below which the atmosphere can be considered to be opaque, and above which is is effectively transparent, from the point of view of its effect on blocking light from a background stellar surface. Again assuming that $h << R_p$, the effective geometric cross section of the planet is then

$$\pi R^2(\lambda) = \pi R_p^2 + 2\pi R_p h(\lambda) \qquad (4)$$

From this we see that the fraction of the star's area that is blocked by the absorbing atmosphere, at a given wavelength, is $f_p(\lambda)$ where

$$f_p(\lambda) = 2\pi R_p h(\lambda) / \pi R_s^2 = 2 R_p h(\lambda)/R_s^2 \qquad (5)$$

The SNR for detecting an atmospheric spectral feature, in an ideal case, is calculated as follows. During a transit, the relevant signal is $N(sig) = N(cont) - N(line)$, where $N(cont)$ is the number of potentially detectable photons in the interpolated continuum at the location of a spectral feature and over the wavelength band of that feature, and $N(line)$ is the number of detected photons in the band. In other words, it is the number of missing photons in the equivalent width of the feature. In terms of the total number of photons detected from the star, in a given spectral range, we have $N(sig) = N(tot)*f_p$. The noise N(noise) is the fluctuation in the total number of detected photons $N(tot)$ in the same wavelength band, so $N(noise) = N^{1/2}(tot)$, ignoring all other noise contributions. Thus the SNR for detecting a given spectral feature is

$$SNR = N^{1/2}(tot) * f_p \qquad (6)$$

An advantage of this formulation is that the $N(tot)$ factor is determined only by the brightness of the star, the bandwidth, and the transit time, while the $f_p$ factor is mostly determined by the atmosphere of the planet.

## 3. MODEL VALIDATION

Our line-by-line radiative transfer code for the Earth has been validated by comparison to observed reflection and emission spectra (Woolf et al. 2002; Turnbull et al. 2006; Christensen & Pearl 1997, Kaltenegger et al. 2007).

To extend the validation data set to rays transiting the atmosphere, especially at low altitudes, we used the solar occultation spectra of the atmosphere as measured from low Earth orbit by the ATMOS 3 experiment (Irion et al. 2002).

model in blue. We show the ATMOS 3 data at the nearest altitude (to within 1 km) to our model, to generate Fig. 2. We did not adjust any parameters in our model for this comparison. A detailed comparison of our model with ATMOS data will be published elsewhere.

Our standard model uses an overall 60% cloud coverage factor for Earth, with the relative proportions of cloud at each altitude being set to be consistent with the Earthshine data (Woolf et al. 2002; Turnbull et al. 2006). The non-overlapping amounts of opaque cloud cover at each altitude are 24% at 1 km, 24% at 6 km, and 12% at 12 km for today's Earth. The effect of shadowing of clouds is taken into account implicitly because the cloud fractions are deduced from modeling the disk-integrated emission spectra of the Earth. A large part of the signal from a disk integrated emission and reflection spectra originates near the limb. Furthermore cloud systems tend to be spaced so that we do not see the cumulative effect of many systems along a given single ray, as we would if a large number of small clouds were scattered uniformly over all of the area. We do not model haze in the atmosphere because our empirical evidence is that in the present atmosphere the optically-thick model cloud layers effectively mimic the effect of optically-thin real haze layers. Fig. 2 shows the variation in atmospheric absorption signatures with height in the atmosphere for selected layers from 0-100 km. We see that the model fits the observed data quite well overall, with some small but systematic departures in the red wing of the 9.6 μm $O_3$ band, and some continuum mismatch (plus and minus) in the 10 km and 12 km rays. These differences are not significant for the present paper.

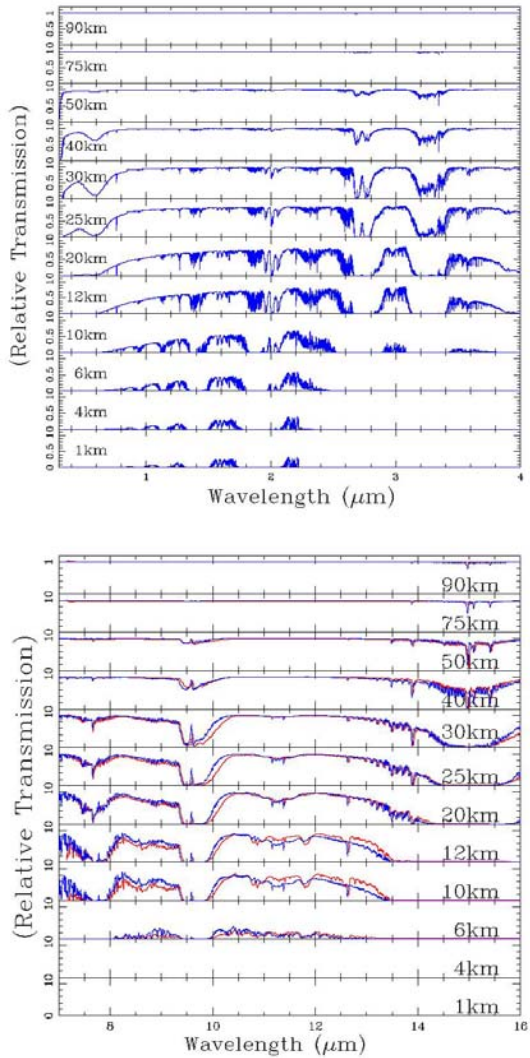

**Fig.2: Transmission spectra of Earth versus tangent height in the clear atmosphere (top: 0.3-4 μm) (right: 4-20 μm). Note that the lower atmosphere can not be probed in transmission for most wavelengths. ATMOS 3 data (Irion et al. 2002) for transmission through the Earth atmosphere (red) and our model (blue) (lower panel)**

We selected a typical set of sunset data[1] from 600-1400 cm$^{-1}$ at geographical latitude of 30.1° (see Fig. 1). The ATMOS 3 data is shown in red and our standard

---
[1] http://remus.jpl.nasa.gov/atmos/atmosversion3/atmosversion3.html

## 4. RESULTS

### 4.1 Transmission Spectrum

Using the model atmosphere and radiative-transfer method sketched above, we calculated the Earth's transmission spectrum, from 0.3-20 μm wavelength, as shown in Figs. 3. Major observable molecular species ($H_2O$, $O_3$, $CH_4$, $CO_2$, $HNO_3$), aerosol absorption, and Rayleigh scattering are labeled. The dark lines show the spectrum smeared to a resolution of $\lambda/\Delta\lambda = 500$ from 0.3-4 μm and $\lambda/\Delta\lambda = 150$ from 4-20 μm, as proposed for JWST (Seager et al. 2009).

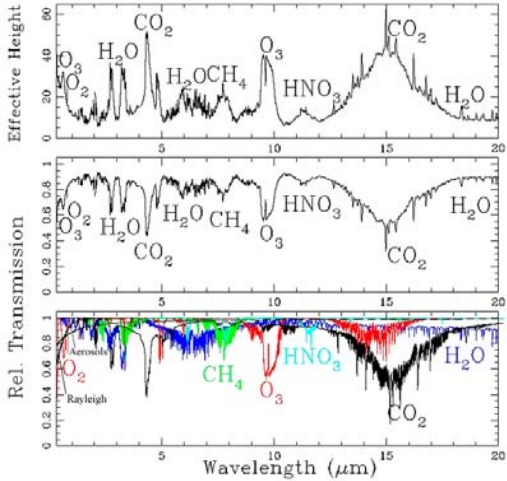

**Fig.3: (top) The effective height of the absorbing atmosphere, for a transiting Earth. Note that the lower 6 km are essentially opaque at all wavelengths, owing mainly to the overlap of line wings plus a small contribution from the continuum opacity of Rayleigh and aerosol scattering. (middle) Transmission spectrum of a 100-km annulus around a transiting Earth. (bottom) The individual species contributions to the total transit spectrum.**

In Fig. 3 the scale on the y axis is relative transmitted intensity $T(\lambda) = \Sigma T_i(\lambda) \Delta h_i / \Sigma \Delta h_i$ in the 0-100 km annular ring. Fig. 3 shows that $O_3$ is relatively strong compared to $H_2O$ in the visible, because the former is located mainly in the clear upper atmosphere, and the latter is located mainly in the lower, more opaque part of the atmosphere. The main detectable features in the 0.3 – 4.0 μm range are $O_3$, $H_2O$, $CO_2$, $CH_4$, and potentially $O_2$, in order of decreasing strength. Detectable features in the infrared transmission spectrum, 4 - 20 μm are $CO_2$, $O_3$, $CH_4$, $H_2O$, and $HNO_3$, in order of decreasing strength.

### 4.2. M-Star Properties for Transits

We calculate the SNR of spectral features of an Earth in the HZ of both a solar-type star and for M stars. The basic physical properties of M stars are given in the first 5 columns of Table 1 (Reid & Hawley 2005). In column 6, the absolute visual magnitude $M_V$ is derived from the *V-I* value and a plot of measured values of $M_V$ vs (V-I) (Reid & Hawley 2005, Reid et al. 1995). In column 7, the semi-major axis in the middle of the habitable zone *a*(HZ, AU), is derived by scaling the Earth-Sun system using $L_{star}/L_{sun} = (R_{star}/R_{sun})^2 (T_{star}/T_{sun})^4$, so $a_{HZ} = 1\ AU\ (L_{star}/L_{Sun})^{0.5}$, and finally

$$a_{HZ} = (T_{star}/5777)^2 (R_{star}/R_{sun}) \qquad (7)$$

This formula assumes that the planet has a similar albedo to Earth, that it rotates or redistributes the insolation as on Earth, and that it has a similar greenhouse effect. Each of these is speculation, but a reasonable starting point given our current lack of knowledge of M-star planets (see Segura et al. 2005, Scalo et al. 2007). We do not adjust the spectrum of the planet in these calculations (see Kaltenegger & Segura en prep). The photon rate from each star is computed assuming that it is a black body, which is a crude approximation for late type stars and leads to an overestimation of the SNR for some of the shorter wavelength transiting spectral features. Column 8 lists the orbital period of a planet $P_{HZ}$ in the *HZ*, and column 9 lists the transit duration $\Delta T_{HZ}$ in hours:

$$P_{HZ} = 365.25*24*a_{HZ}^{3/2}M_{star}^{-1/2} \qquad (8)$$

$$\Delta T_{HZ} = P_{HZ}*(2R_{star})/(2\pi a_{HZ}) \qquad (9)$$

Here $a_{HZ}$ is given in AU, the mass $M_{star}$ and the radius of the star $R_{star}$ in solar units. Typical transit times are on the order of an hour. Column 10 lists the ratio $\Delta I/I$ of the depth of the overall transit signal $\Delta I$ to the non-transit signal I.

$$\Delta I/I = (R_p(\lambda)/R_s)^2 \qquad (10)$$

### 4.3 SNR and Statistics of Transits

We calculate the achievable SNR for primary eclipse measurements. The value of *N(tot)* is calculated assuming an effective temperature of 5770 K for the Sun and the values given in Table 1 for M stars. The number of detected photons is computed assuming a 6.5-m diameter telescope (like JWST) in space, a net efficiency of 0.15 electrons/photon, and an integration time equal to the transit time.

Table 2 lists the strongest features from the effective height spectrum in Fig. 3, including central wavelength, full width at half maximum, and the average effective height of the feature. Column 5 gives the SNR for each feature for current Earth in the HZ of the Sun, and columns 6 – 10 give the SNR values for these features for the current Earth in the HZ of M0 – M9 dwarf stars, all for a 6.5-m telescope in space, with an efficiency of 0.15 electrons/photon, for a single transit, and at a standard distance of 10 pc.

Noting that nearly all of the SNR values in Table 2 are less than unity, it is of interest to ask what SNR values could be achieved if the integration time was not a single transit, but instead a fixed amount of telescope time. Values of SNR for 200 hrs of co-added transit time are listed in Table 3, again for a Sun-like star and M stars, all at 10 pc. Details on the 200-hrs cases are given in Table 4, which lists the number of transits per year, the number of transits in a total observing time of 200 hrs (where the telescope only observes during the transit itself), and the number of calendar years needed to achieve all 200 hrs.

The closest star to us, in each category, is of interest. For G2V stars, the closest one is α Cen A at 1.3 pc. For the M stars, we list the nearest one in each sub-class in Table 5, from the list of nearby M stars by N. Reid (in prep). All of these closest stars are nearer than 5 pc, so if there is a transiting Earth around one of these, the observations can be done more efficiently than was assumed for the 10 pc cases.

## 5. DISCUSSION

### 5.1 Apparent Radius versus Wavelength

The inclusion of realistic cloud coverage has a surprisingly small effect on the apparent radius or the strengths the spectral features. For models with and without clouds, we find that the rms difference in effective height is negligibly small, about 0.1 km rms over the whole spectrum. This is because the Earth's atmosphere, even without clouds, shows low transmission in the lower atmosphere layers (see Fig. 3 for details). Rayleigh scattering and aerosol absorption have a strong effect for the UV-visible wavelength range, and line wings in the infrared range, because they block most light in the lower part of the atmosphere.

The apparent radius of the Earth in transmission between 0.3-20 µm varies between about 6-50 km above ground. It is determined mostly by water and carbon dioxide absorption for 2.5-20 µm, with an added component due to aerosol absorption from 0.6-2.5 µm, and added Rayleigh scattering and ozone absorption from 0.3-0.6 µm. The Rayleigh scattering is a proposed way to detect the most abundant atmospheric species (probably $H_2$ for an EGP, and $N_2$ for Earth-like planets). The Earth appears to be 50 km bigger in the UV than its solid-body radius. This simple test shows that especially for planets with a denser atmosphere than Earth, we will not be able to probe the lower atmosphere. Even for Earth the lowest 6 km above ground is not accessible in any region between 0.3-20 µm. In terms of error on the deduced radius, this is not a big concern, but it severely influences the detectability of atmospheric signatures and potential biosignatures in certain wavelength ranges (e.g., water in the visible wavelength range). Fig. 3 shows the effect of individual components in the atmosphere as well as Rayleigh scattering and aerosol absorption on the apparent radius of the planet.

5.2 Spectral Features

The simulations show that only ozone and potentially oxygen are detectable in the UV-visible in Earth's transmission spectra. The near-IR shows absorption features of $CO_2$, $H_2O$ and potentially $CH_4$. The mid-IR shows absorption features of $CO_2$, $H_2O$, $O_3$, $CH_4$ and $HNO_3$.

The strong absorption features of oxygen and ozone in the visible are reduced by the effect of aerosol absorption and Rayleigh scattering on the overall spectrum. Ozone has strong absorption bands, the Hartley and Chappuis band (200–350 nm and 420–830 nm respectively), and the absorption feature at 9.6 µm in the infrared that can easily be seen in the transmission spectrum (see Fig. 3). Molecular oxygen has several bands, the strongest $O_2$ feature is the saturated Fraunhofer A-band at 0.76 µm, weaker feature are at 0.7 µm and 1.26 µm. The other molecular oxygen features are overlapping with $CO_2$ and $H_2O$ absorption. Methane can be detected in the mid-IR at 7.66 µm and potentially in the near-IR at 2.4 µm. Carbon-dioxide has an extremely strong absorption feature at 15 µm with extended wings, and weaker detectable features at 4.5 µm, 2.75 µm and 2 µm. One can see increasingly strong and detectable $H_2O$ bands from 1.14 µm, 3.3 µm, 6.3 µm to its rotational band that extends from 12 µm out into the microwave region and its extended wings. $HNO_3$ can also be detected in the mid-IR at 11.5 µm.

5.3 SNR for Primary Transits

The resulting SNR values for a 6.5-m space-telescope like JWST for one transit are all small, with most less than unity (see Table 2) except for the closest stars (see Table 6). A factor that can be used to increase the SNR for the M-star case is that multiple transits can be expected, given the relatively short period of a planet in the HZ; the number of transits per year ranges from about 6 - 192 for M0V to M9V. The SNR can be increased by the square root of these factors, i.e., by about 2 - 13 times, which increases the net SNR values especially for late type stars. For an observation time of 200 hrs of co-added transmission data high SNRs can be achieved and are given in Table 3 (200 hrs was chosen for a comparison to Seager et al. 2009). Table 4 shows the number of transits that need to be co-added to achieve a 200 hrs transit signal, 37 to 469 for M0V to M9V respectively.

The SNR values scale inversely as the distance to the star, so if closer examples are found, the SNR will increase proportionately, making close stars the best targets. The SNR values also scale linearly as the diameter of the telescope, so a 15-m and a 40-m telescope improves these values by a factor of 2.3 and 6.1 respectively. For a 15-m space telescope the SNR for an Earth transiting a Sun at 10 pc is between 0.4 and 3.8, for a 40-m space telescope the SNR is between 1.2 and 10.3. As pointed out by Selsis in 2004, a telescope that could detect any transmission features on potentially habitable planets in the visible during one transit would have to have a minimum diameter of 30-m to 40-m and have to be operated in space (Ehrenreich et al. 2006). An equivalent collection area could detect features in emission and reflected light that characterize the planet itself in more detail (Des Marais et al. 2002; Kaltenegger & Selsis 2007).

### 5.4 Implication for Transit Search

These calculations show that the stability of the instrument is crucial to probe extrasolar Earth-like planets for biomarkers in transmission, due to the small scale height and extent of Earth's atmosphere – and therefore also the volume of the small ring that can be probed for biomarkers in transit. Past calculations have found encouraging results for atmospheric absorption feature detection by applying the transit technique to Mars-sized planets with a 1 bar atmosphere, and a much larger molecular column density than Earth, which improves the SNR – as discussed below. To detect biomarkers one would ideally study a planet with a large scale height as well as an extremely extended atmosphere. Such an ideal planet has been envisioned (Ehrenreich et al. 2006) and used to calculate observation scenarios for JWST (Seager et al. 2009). The planet was designed by decreasing Earth's mass to 10% (like Mars), so with 50% of Earth's radius, and imposing a 1 bar surface pressure. Assuming that such a planet could maintain an extended Earth-analog atmosphere, this would result in an atmospheric height of 260 km, with a scale height of 24 km instead of Earth's 8.8 km (Ehrenreich et al. 2006). Such a Mars size planet, that would maintain an extended 260 km Earth-analog atmosphere, has been previously called 'Small-Earth'. Whether such a planet could maintain its atmosphere - unlike Mars - and could exist, is not clear, but it would be an ideal target for transit spectroscopy, better than Earth. The SNR calculated in Table 2 would improve by about a factor 2.5. That factor is generated by moving from a scale height of 8.8 km on Earth to a 22 km scale height on such an ideal planet (which is the effect of decreased gravity on an ideal planet with half of Earth's radius):

$$f_{p2}/f_{p1} = (2R_p \, 2.5h(\lambda))/(2R_p h(\lambda)) \cong 2.5$$

In addition an integration time of 200 hrs for transiting planets was assumed in this scenario (Seager et al. 2009), which translates to 37 to 469 transits and observations over a time span of about 6.7 to 2.5 years for M0V to M9V respectively (see Table 4). The capacity to co-add transits will be a crucial requirement for an atmospheric transit search for telescopes like JWST.

Note that for Earth the results by Ehrenreich et al. 2006 (0.3 to 2.0 μm) correspond to our results – if no dust or aerosol absorptions are included in our model, a complete cloud cover is set at 10 km, with a slight overestimation of the $CO_2$ features in the Ehrenreich paper. For a comparison of results note that Table 3 in Ehrenreich et al. 2006 uses a telescope effective mirror size of 10 m, which translates to a telescope diameter of about 67 m for an efficiency of 0.15.

### 5.5 Estimating the Distance of the Most Likely Transiting Star

To estimate the distance to the nearest likely transited star, we use $d_n = (3n/4\pi\rho)^{1/3}$, where the nearest transiting star is the number n. This is also the average number of stars one needs to follow before finding a transit. Here $\rho$ is the space density of stars of the class under consideration, $n = 1/p$, where p is the probability of a transit, and $p = R_s/a_p$, with $R_s$ the radius of the star and $a_p$ the semi-major axis of the planet (Borucki & Summers 1984).

For GV stars we use $\rho \sim 0.005$ stars/pc$^3$, R(star) ~ 1 R(sun), and a(planet) ~ 1 AU, so the nearest likely case of a planet in the HZ transiting its GV star, assuming that all such stars have such a planet, is number n(GV) ~ 216, and this star will be at about 22 pc.

For M stars we use the M-star count, complete within 8 pc of the Sun, by N. Reid, as represented by his list of 118 such stars (Reid et al. in prep). About half of the stars in this list are class M3 and M4, and the other half is split between earlier and later types. From this 8pc list we estimate of the space density of each type. Combining this with the above data on star properties, we derive estimates of the distance to the nearest likely HZ planet transiting an M star. These values are roughly 20 pc for very early M stars, dropping to about 10 pc for M3 and M4 stars, and rising again to about 20 pc for very late types. The median likely distance of an M star with a transiting planet in the HZ, assuming all stars have such a planet, is about 13 pc. This value is roughly the same as the 10 pc of Table 2, thereby making Table 2, representative of what we might expect to find. .

## 6. CONCLUSIONS

Concentrating on Earth, we calculated the expected transit spectrum. We then placed this Earth-analog planet in the habitable zone of a Sun-like, as well as several M stars and calculated the ideal SNR values only considering photon noise from the star as a noise source for the strongest absorption features in the spectrum, from the ultraviolet to the thermal infrared, for transits at 10 pc distance, as well as the closest G2V and M0V to M9V stars, assuming a 6.5-m space based telescope (like JWST).

In the transit spectrum in the UV-visible, only $O_3$ and potentially $O_2$ are detectable. The near-IR shows absorption features of $CO_2$, $H_2O$ and potentially $CH_4$. The mid-IR shows absorption features of $CO_2$, $H_2O$, $O_3$, $CH_4$ and $HNO_3$. For M stars the features in the IR are easier detectable than the features in the visible due to the peak of the stellar flux in the infrared. Most of the lower atmosphere can not be probed in transmission, mainly due to Rayleigh scattering and aerosol absorption in the UV-visible and $H_2O$ and $CO_2$ absorption in the IR (see Fig. 3). The apparent radius of the Earth varies by a maximum of 50 km due to absorption, less than 1%. The SNR values per transit are all small, on the order of unity or less (except for the closest stars).

We extended the calculation to the cases of larger telescopes and multiple transits. Our calculations show that multiple transits are needed to detect atmospheric features on an Earth-analog in transit. Table 2 shows that a G2V star gives a better SNR per transit in the visible than a small M star because of the short transit duration of a planet orbiting an M star, but about the same and better in the mid-infrared. Assuming a set of co-added multiple transits, the smallest M stars (M9V and M8V) are the best targets for spectroscopy in transit in the near-IR to IR because of the favorable contrast of the star to the planet (see Table 3). Stability of the instrument and a method to co-add transit observations will be a crucial requirement for atmospheric transit search for telescopes like JWST.

**Acknowledgement:**
We are grateful to Neill Reid for providing us with the M-star data and Todd Henry for providing us with a list of the closest M stars. Special thanks to Ken Jucks, Sara Seager and Phillip Nutzman for stimulating discussion and comments. This work was sponsored by NASA grant NAG5-13045 at the Harvard Smithsonian Center for Astrophysics and the Origins of Life Initiative at Harvard. Part of the research described in this paper was carried out at the Jet Propulsion Laboratory, California Institute of Technology, under a contract with the National Aeronautics and Space Administration.

**Table 1. The physical characteristics of each sub-class of M stars are listed here, along with derived values related to an Earth-size planet in the HZ.**

| SpTy dwarf | T K | R $R_{sun}$ | Mass $M_{sun}$ | L/100 $L_{sun}$ | $M_V$ Mag | a(HZ) AU | P(HZ) hr | ΔT(HZ) hr | ΔI/I % |
|---|---|---|---|---|---|---|---|---|---|
| M0 | 3800 | 0.62 | 0.60 | 7.2 | 9.34 | 0.268 | 1571 | 5.37 | 0.022 |
| M1 | 3600 | 0.49 | 0.49 | 3.5 | 9.65 | 0.190 | 1039 | 3.96 | 0.035 |
| M2 | 3400 | 0.44 | 0.44 | 2.3 | 10.12 | 0.152 | 786 | 3.36 | 0.043 |
| M3 | 3250 | 0.39 | 0.36 | 1.5 | 11.15 | 0.123 | 633 | 2.96 | 0.055 |
| M4 | 3100 | 0.26[2] | 0.20 | 0.55 | 12.13 | 0.075 | 401 | 2.06 | 0.124 |
| M5 | 2800 | 0.20 | 0.14 | 0.22 | 16.0 | 0.047 | 238 | 1.50 | 0.209 |
| M6 | 2600 | 0.15 | 0.10 | 0.09 | 16.6 | 0.030 | 147 | 1.07 | 0.372 |
| M7 | 2500 | 0.12 | ~0.09 | 0.05 | 18.8 | 0.022 | 98 | 0.78 | 0.582 |
| M8 | 2400 | 0.11 | ~0.08 | 0.03 | 19.8 | 0.019 | 81 | 0.69 | 0.69 |
| M9 | 2300 | 0.08 | ~0.075 | 0.015 | 17.4 | 0.013 | 46 | 0.43 | 1.31 |

**Table 2: Major spectroscopic features (col. 1-4) and SNR (col. 5-10) of a transiting Earth per transit, for a 6.5-m space based telescope, for the Sun and M stars at 10 pc.**

| | 6.5-m telescope | | | SNR (E, Star) 10 pc | | | | | |
|---|---|---|---|---|---|---|---|---|---|
| Feature | λ(μm) | Δλ(μm) | H(λ),km | G2V | M0V | M2V | M5V | M8V | M9V |
| $O_3$ | 0.6 | 0.15 | 10 | 1.67 | 0.58 | 0.45 | 0.32 | 0.20 | 0.17 |
| $H_2O$ | 1.9 | 0.2 | 5 | 0.47 | 0.32 | 0.31 | 0.35 | 0.34 | 0.35 |
| $CO_2$ | 2.8 | 0.1 | 20 | 0.84 | 0.62 | 0.62 | 0.74 | 0.77 | 0.79 |
| $H_2O$ | 3.3 | 0.25 | 20 | 1.08 | 0.81 | 0.82 | 1.01 | 1.06 | 1.10 |
| $CH_4$ | 7.7 | 0.7 | 7 | 0.20 | 0.16 | 0.17 | 0.22 | 0.24 | 0.25 |
| $O_3$ | 9.8 | 0.7 | 30 | 0.61 | 0.50 | 0.52 | 0.67 | 0.75 | 0.79 |
| $CO_2$ | 15.2 | 3.0 | 25 | 0.58 | 0.48 | 0.50 | 0.65 | 0.74 | 0.78 |

**Table 3: Major spectroscopic features (col. 1) and SNR (col. 2-12) of a transiting Earth for a total co-added observation time of 200 hrs, for a 6.5-m space based telescope for the Sun and M stars**

| Feature | G2V | M0V | M1V | M2V | M3V | M4V | M5V | M6V | M7V | M8V | M9V |
|---|---|---|---|---|---|---|---|---|---|---|---|
| $O_3$ | 16.9 | 9.1 | 9.7 | 8.9 | 8.6 | 9.2 | 9.4 | 9.5 | 9.6 | 8.6 | 9.6 |
| $H2O$ | 4.8 | 5.0 | 6.0 | 6.2 | 6.6 | 7.9 | 10.5 | 13.0 | 14.7 | 14.9 | 18.9 |
| $CO2$ | 8.5 | 9.7 | 11.7 | 12.3 | 13.3 | 16.1 | 22.2 | 28.2 | 32.5 | 33.7 | 43.4 |
| $H_2O$ | 11.0 | 12.8 | 15.5 | 16.4 | 17.7 | 21.6 | 30.1 | 38.5 | 44.6 | 46.4 | 60.2 |
| $CH_4$ | 2.0 | 2.5 | 3.1 | 3.3 | 3.6 | 4.5 | 6.5 | 8.5 | 9.9 | 10.5 | 13.8 |
| $O_3$ | 6.2 | 7.8 | 9.5 | 10.3 | 11.2 | 13.9 | 20.0 | 26.3 | 30.9 | 32.7 | 43.2 |
| $CO_2$ | 5.9 | 7.5 | 9.2 | 9.9 | 10.9 | 13.5 | 19.5 | 25.8 | 30.4 | 32.2 | 42.6 |

---

[2] note that the original references (Reid et al 2005) states 0.36, but this should be 0.26 as shown.

Table 4. Transits of G2V and M stars per Earth-year (col. 1). number of transits to add up to 200 hrs (col. 2), and number of years to accumulate 200 hrs of transit observation time for an Earth-size planet in the HZ of its host star (col.3).

| SpTy dwarf | # transits Per year | 200 hrs # transits | 200 hrs # years |
|---|---|---|---|
| G2 | 1.0 | 15.4 | 15.4 |
| M0 | 5.6 | 37.2 | 6.7 |
| M1 | 8.4 | 50.5 | 6.0 |
| M2 | 11.1 | 59.6 | 5.3 |
| M3 | 13.8 | 67.6 | 4.9 |
| M4 | 21.8 | 97.0 | 4.4 |
| M5 | 36.7 | 133.2 | 3.6 |
| M6 | 59.7 | 186.6 | 3.1 |
| M7 | 89.1 | 257.3 | 2.9 |
| M8 | 108.1 | 287.9 | 2.7 |
| M9 | 191.8 | 469.0 | 2.5 |

Table 5. The closest stars for each M substellar class from Reid et al. (in prep).

| Name | d(pc) | Sp Type |
|---|---|---|
| Gl 887 | 3.29 | M0.5 |
| Gl 15 A | 3.56 | M1 |
| Gl 411 | 2.54 | M2 |
| Gl 729 | 2.97 | M3.5 |
| Gl 699 | 1.83 | M4 |
| Gl 551 | 1.30 | M5.5 |
| Gl 406 | 2.39 | M6 |
| Gl 473 B | 4.39 | M7 |
| SCR1845-63A | 3.85 | M8.5 |
| Denis1048 | 4.03 | M9 |

Table 6: Major spectroscopic features and SNR of a transiting Earth for a single transit, for a 6.5-m space based telescope, for the closest stars per stellar subtype (see Table 5). For G2V, we use α centauri A, Gl 559 A at 1.34pc.

| Feature | G2V | M0V | M1V | M2V | M3V | M4V | M5V | M6V | M7V | M8V | M9V |
|---|---|---|---|---|---|---|---|---|---|---|---|
| $O_3$ | 12.5 | 1.8 | 1.5 | 1.8 | 1.4 | 2.0 | 2.4 | 1.1 | 0.5 | 0.5 | 0.4 |
| $H_2O$ | 3.5 | 1.0 | 0.9 | 1.2 | 1.0 | 1.7 | 2.7 | 1.5 | 0.8 | 0.9 | 0.8 |
| $CO_2$ | 6.3 | 1.9 | 1.8 | 2.4 | 2.1 | 3.5 | 5.8 | 3.3 | 1.8 | 2.0 | 1.9 |
| $H_2O$ | 8.1 | 2.5 | 2.4 | 3.2 | 2.8 | 4.7 | 7.8 | 4.4 | 2.4 | 2.8 | 2.7 |
| $CH_4$ | 1.5 | 0.5 | 0.5 | 0.7 | 0.6 | 1.0 | 1.7 | 1.0 | 0.5 | 0.6 | 0.6 |
| $O_3$ | 4.5 | 1.5 | 1.5 | 2.0 | 1.8 | 3.0 | 5.2 | 3.0 | 1.7 | 1.9 | 1.9 |
| $CO_2$ | 4.3 | 1.4 | 1.4 | 2.0 | 1.7 | 2.9 | 5.0 | 3.0 | 1.7 | 1.9 | 1.9 |